\documentclass[9pt,twocolumn,twoside]{opticajnl}
\journal{opticajournal} 
\usepackage{siunitx}
\setboolean{shortarticle}{true}
\usepackage{lineno}
\usepackage{pdfpages}

\title{Inverted pin beams for robust long-range propagation through atmospheric turbulence}

\author[1,2,*]{Sotiris Droulias}
\author[3]{Michalis Loulakis}
\author[3,4]{Dimitris G. Papazoglou}
\author[3,4]{Stelios Tzortzakis}
\author[5,6]{Zhigang Chen}
\author[1,5,7,**]{Nikolaos K. Efremidis}

\affil[1]{Institute of Applied and Computational Mathematics, FORTH, 70013 Heraklion, Crete, Greece}
\affil[2]{Department of Digital Systems, University of Piraeus, Piraeus 18534, Greece}
\affil[3]{Institute of Electronic Structure and Laser, Foundation for Research and Technology - Hellas, P.O. Box 1527, 71110 Heraklion, Crete, Greece}
\affil[4]{Department of Materials Science and Technology, University of Crete, 700 13 Heraklion, Crete, Greece}
\affil[5]{MOE Key Laboratory of Weak-Light Nonlinear Photonics, TEDA Applied Physics Institute and School of Physics, Nankai University, Tianjin 300457, China}
\affil[6]{Collaborative Innovation Center of Extreme Optics, Shanxi University, Taiyuan, Shanxi 030006, China}
\affil[7]{Department of Applied Mathematics, University of Crete, Heraklion 71409, Greece}
\affil[*]{sdroulias@unipi.gr}
\affil[**]{nefrem@uoc.gr}

\begin{abstract}
  We introduce a new class of optical beams, which feature a spatial profile akin to an ``inverted pin''. In particular, we asymptotically find that close to the axis the transverse amplitude profile of such beams takes the form of a Bessel function with width that gradually increases during propagation. We examine numerically the behavior of such inverted pin beams in turbulent environments as measured via the scintillation index, and show that they outperform Gaussian beams (collimated and focused) as well as Bessel beams and regular pin beams, which are all optimized, especially in the moderate and strong fluctuation regimes.
\end{abstract}

\setboolean{displaycopyright}{false} 

\begin{document}

\maketitle

Gaussian beams, are the ``natural'' modes of most lasers. They are subjected to beam spreading due to diffraction, and, owing to the simplicity in their generation, they are most commonly used in a variety of applications. However, there are cases where specially engineered classes of beams can have better performance in specific applications. These include propagation-invariant beams, such as Bessel beams~\cite{Durnin1987a, Durnin1987b, McGloin2010} and Airy beams and accelerating waves~\cite{Siviloglou2007a, Siviloglou2007b, efrem-optica2019}.
Bessel-like beams have a profile that resembles a Bessel function close to the axis, with an engineered width and amplitude~\cite{Cizmar2009, gouts-ol2020} or even trajectory~\cite{chrem-ol2012-bessel, zhao-ol2013}. Along these lines, a family of the so-called pin beams with a width that gradually decreases during propagation and a Bessel-like profile close to the axis has been proposed and demonstrated~\cite{Zhang2019,Li2020,bongi-pr2021}.

Beam propagation through the atmosphere has been the subject of numerous theoretical and experimental studies, with several applications in astronomical imaging, remote sensing, laser radar, and free-space optical communications~\cite{Andrews2005}. The main challenge is that beam propagation in the atmosphere rarely occurs unperturbed; small particles tend to attenuate the beam via absorption and scattering. Importantly, temperature and pressure variations give rise to atmospheric turbulence that causes fluctuations in the refractive index and, as a result, to the beam irradiance. While the refractive index fluctuations are relatively weak (of the order of $10^{-6}$--$10^{-8}$), propagation for several hundred meters leads to cumulative effects that may severely distort the beam wavefront. 
Fluctuations of the beam irradiance, commonly described as ``scintillations'', have been extensively studied for several types of beams under turbulent conditions. Early studies of scintillation were mainly focused on Gaussian beams~\cite{Miller1993, Miller1994}. The use of structured light beams as a means to reduce laser beam scintillations has recently started to attract attention. In this respect, Bessel beams~\cite{eyyub-apb2007, Eyyuboglu2009, eyyub-ao2013}, Airy beams~\cite{Gu2010a, chu-ol2011, Ji2013, Nelson2014}, and pin beams~\cite{Zhang2019,hu-ol2022,cao-josaa2022} have been utilized.

In this work, we introduce a new class of structured light beams that is characterized by a main lobe with a Bessel-like transverse profile and a width that gradually increases during propagation. Due to their shape, we call such beams inverted pin beams (IPBs). Using asymptotic calculations, we first analyze the generation and dynamics of IPBs. We numerically show that IPBs have a significantly reduced scintillation index when compared to four other classes of beams in atmospheric turbulence. In particular, IPBs are compared to collimated Gaussian beams (CGBs), focused Gaussian beams (FGB), pin beams (PBs), and Bessel beams (BBs) which are all optimized for the same conditions. Importantly, we show that the IPBs optimized for a specific set of parameters, can have a very small scintillation index for a wide range of turbulent conditions and propagation distances. We expect our proposed beams might find applications in free-space optical communications and optical imaging.

\begin{figure}[t!]
  \centering
  \includegraphics[width=\linewidth]{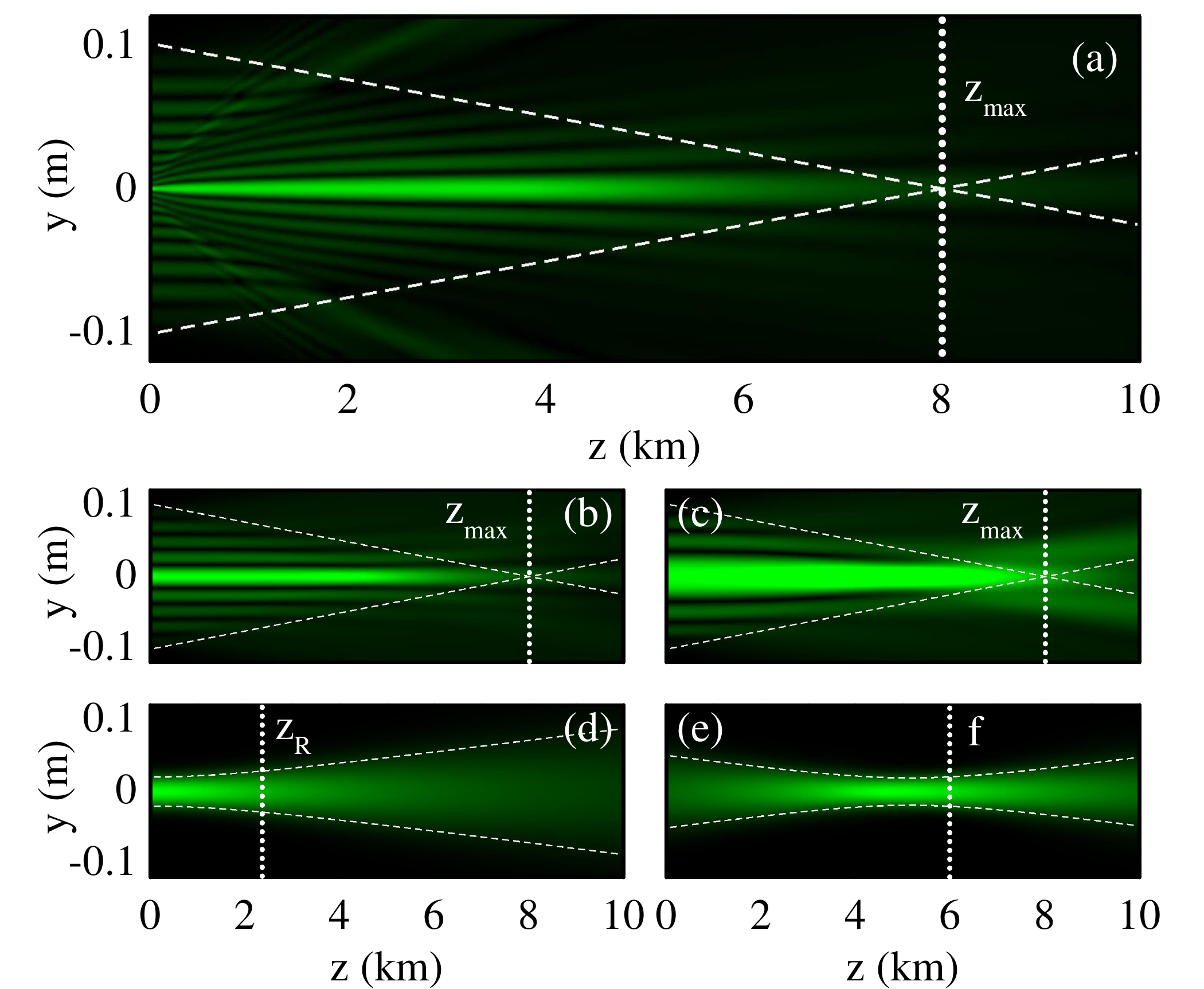}  
  \caption{Field amplitude dynamics of (a) an IPB with $\gamma=0.4$ and comparison with (b) a PB with $\gamma=1$ (or BB), (c) a PB with $\gamma=1.4$, (d) a CGB and (e) a FGB.
    In (a), (b), (c) $z_\mathrm{max}=8$ \unit{\kilo\meter} is the maximum propagation distance of these beams, and the dashed lines depict the cone generated by the outer rays
    (starting at $\rho=D_t/2$).
    In (d), the width of the CGB is $w_0=20$ \unit{\milli\meter} corresponding to a Rayleigh length $z_R\approx2.4$ \unit{\kilo\meter} (the focal length is $f=\infty$), whereas in (e) the FGB has $w_0=50$ \unit{\milli\meter} and $f=6$ \unit{\kilo\meter}. In (d), (e) the dashed curves mark the evolution of the beam diameter.}
  \label{fig:01}
\end{figure}

We consider a beam propagating along the $z$-direction, and denote the radial coordinate in the transverse $(x,y)$ plane as $r=\sqrt{x^2+y^2}$.
We define the radial coordinate $\rho$ on the initial plane $z=0$ and assume an input beam profile
\begin{equation}
  \psi_0(\rho) = A(\rho)e^{i\phi(\rho)},
  \label{eq:psi0}
\end{equation}
where $A(\rho)$ is the amplitude, and the phase has a radial power-law dependence $\phi(\rho) = -kC\rho^\gamma$. Also, $k=2\pi/\lambda$ is the wavenumber, $\gamma$ is the power-law exponent, and $C$ determines the speed of the phase variations. Under the assumption of a slowly varying amplitude $A(\rho)$, and following the analysis of~\cite{gouts-ol2020,Li2020}, asymptotic calculations lead to the following relation for the beam profile
\begin{equation}
  \psi(r, z)=\sqrt{\frac{2 \pi k}{2-\gamma}}
  \left(C \gamma z^{\frac{\gamma}{2}}\right)^{\frac{1}{2-\gamma}} A(\rho(z))
  J_0\left(\frac{k r\rho(z)}{z}\right)
  e^{i\Phi},
  \label{eq:psirz}
\end{equation}
where
\begin{equation}
  \rho(z) = (C\gamma z)^{1/(2-\gamma)}.
  \label{eq:rhoz}
\end{equation}
and
$
\Phi =
k r^2/(2z)+(C^2 \gamma^2 z^\gamma)^{1/(2-\gamma)} (k/2)(1-2/\gamma)-\pi/4.
$
In terms of ray optics, Eq.~(\ref{eq:rhoz}) relates the radial displacement $\rho$ of a ray on the input plane to its focal distance $z$. As a result, if we want to generate a beam that maintains its Bessel-like shape up to a maximum propagation distance $z_\mathrm{max}$ with a transmitter aperture diameter $D_t$, then from Eq.~(\ref{eq:rhoz}) we find that
\begin{equation}
  C(D_t,z_\mathrm{max}) =
  [1/(\gamma z_\mathrm{max})]
  (D_t/2)^{2-\gamma}.
  \label{eq:CDtzmax}
\end{equation}

We can derive a simplified real profile for the initial waveform $\psi(\rho,z=0)$ using Bessel functions. In particular, we express the initial beam amplitude as
\begin{equation}
  A(\rho) = B(\rho) \rho^{-\gamma/2}
  \sqrt{
    (2-\gamma)/
    (2\pi k\gamma C)
  }.
  \label{eq:Arho}
\end{equation}
We then add in Eq.~(\ref{eq:psi0}) its complex conjugate to obtain a real function. The rays of the complex conjugate are diverging (moving away from the axis) and, thus, do not affect the beam dynamics close to the optical axis. Using large argument Bessel asymptotics, the initial condition can be expressed as
\begin{equation}
  \psi_0(\rho) =
  B(\rho) 
  J_0(kC\rho^\gamma).
  \label{eq:psi02}
\end{equation}
The initial profile of Eq.~(\ref{eq:psi02}) takes the form of a Bessel function with a generic power-law dependence from the radius. 
From Eq.~(\ref{eq:Arho}) the asymptotic formula of Eq.~(\ref{eq:psirz}) is simplified to
\begin{equation}
  \psi(r,z) =
  B(\rho(z)) 
  \sqrt{
    \frac{
      2-\gamma
    }{
      \gamma 
    }
  }
  J_0\left(
    \frac{kr\rho(z)}{z}
  \right).
  \label{eq:psirz2}
\end{equation}
We note that for constant $B(\rho)$, independently of the value of $\gamma$, the on-axis amplitude of the beam remains constant during propagation. From the argument of the Bessel function in Eq.~(\ref{eq:psirz2}), we see that the full width at half maximum of the beam is 
\begin{equation}
  \operatorname{FWHM}(z) = 2.2527 k^{-1} (C\gamma z^{\gamma-1})^{\frac{1}{\gamma-2}}.
  \label{eq:FWHM}
\end{equation}

Importantly, for different values of $\gamma$, the generated beams exhibit significantly different behavior. In particular, for $1<\gamma<2$, a regime explored in~\cite{Zhang2019, Li2020}, the width of the beam decreases during propagation leading to a pin resembling beam. On the other hand, for $\gamma=1$ the width of the beam remains constant independently of $z$ -- a Bessel beam~\cite{Durnin1987a, Durnin1987b}. Finally, for $0<\gamma<1$ the width of the beam increases during propagation resulting to an inverted pin beam profile [see Fig.~\ref{fig:01}(a)]. This regime will be explored in this work.

The behavior of IPBs is also compared with Gaussian beams with input profile
\begin{equation}
  \psi_0(\rho)=A\exp(-\rho^2/w_0^2-i k\rho^2/(2f)),
  \label{eq:Eq1}
\end{equation}
where $w_0$ is the beam radius on the input plane and $f$ the focal distance. For $f\rightarrow{\infty}$, the beam is collimated and the only tuning parameter is the input beam radius $w_0$. The closed-form beam dynamics of Eq.~(\ref{eq:Eq1}) can be found for example in~\cite{Andrews2005}.

In this paper, we select a set of parameters that is similar to our previous works~\cite{gouts-ol2020,Li2020}: In particular, the optical wavelength is $\lambda=532$ \unit{\nano\meter}, the aperture diameter of the transmitter is $D_t=20$ \unit{\centi\meter}, whereas $\gamma=0.4$ for IPBs and $\gamma=1.4$ for PBs.

In Fig.~\ref{fig:01}, we compare the dynamics of IPB with BB, PB, CGB, and FGB in the absence of turbulence. As we will see in the next sections their parameters
are all selected to minimize the scintillation index for long-range $L=6$ km of propagation. All structured light beams exhibit self-similar Bessel-like behavior close to the axis. On the other hand, the width of the CGB increases during propagation due to diffraction, whereas the FGB has a minimum width close to the focal plane. Importantly, the IPB has the smallest main lobe as compared to all of the other classes of beams everywhere through the range $\numrange{1}{6}$ \unit{\kilo\meter}.

\begin{figure}[t!]
  \centering
  \includegraphics[width=\linewidth]{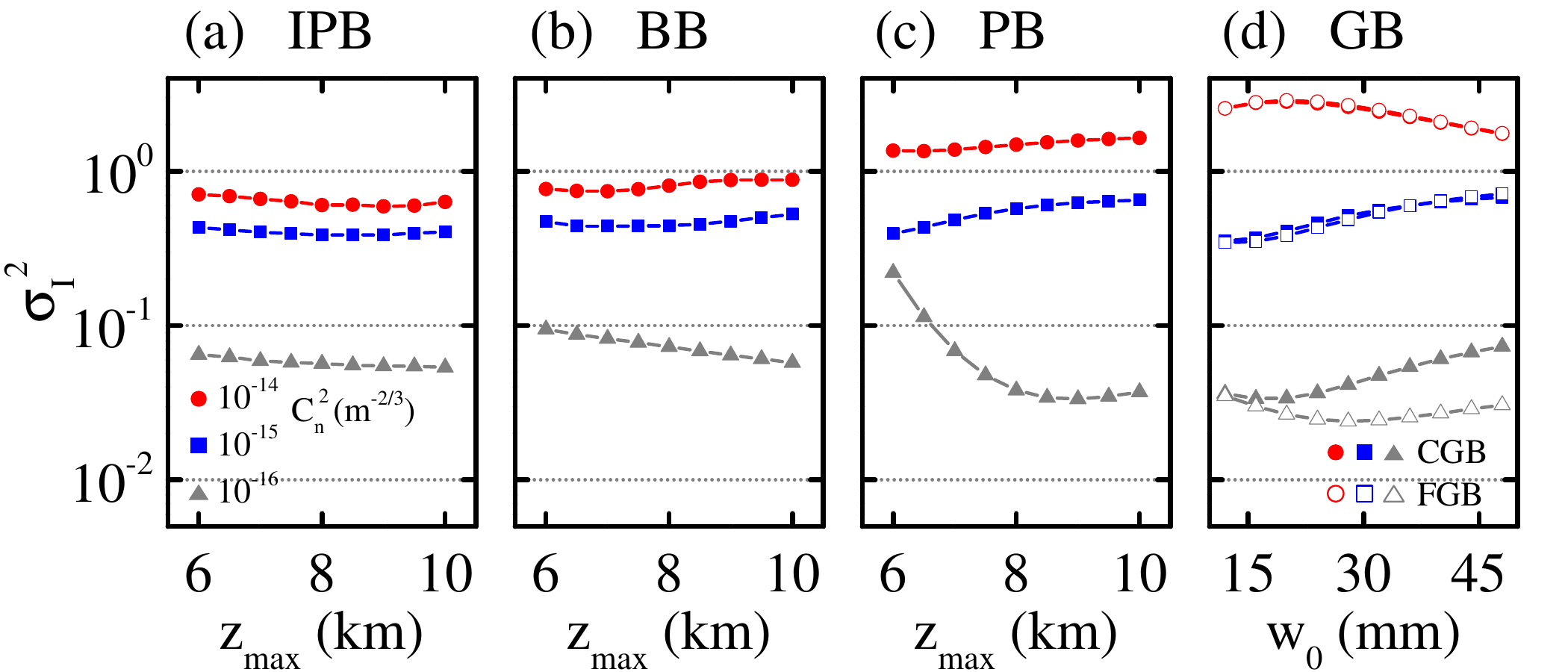}  
  \caption{Diagrams used for optimal beam selection at receiver distance $L=6$ \unit{\kilo\meter}, based on the scintillation index for different turbulence strengths ($C_n^2=10^{-16},10^{-15},10^{-14}$ \unit{\meter^{-2/3}}). In particular, in (a)-(c) we depict $\sigma_I^2$ vs. $z_\mathrm{max}$ for IPB, BB, and PB, respectively. In (d) we display $\sigma_I^2$ vs. $w_0$ for CGB and FGB. The beam parameters are the same as in Fig.~\ref{fig:01}.}
  \label{fig:02}
\end{figure}

Our main goal is to compare the performance of the beams shown in Fig.~\ref{fig:01}, under different turbulent strengths $C_n^2$. In this respect, the first task is to separately optimize each one of these beams in terms of the scintillation index $\sigma_I^2=\langle I^2\rangle/\langle I\rangle^2-1$, where $I=|\psi|^2$ is the beam intensity and the brackets denote ensemble average. 
Note that the refractive index fluctuations in atmospheric turbulence are characterized by the refractive index structure parameter $C_n^2$.  Typically, for weak turbulence $C_n^2\sim10^{-17}$ \unit{\meter^{-2/3}} or less, whereas for strong turbulence $C_n^2\sim10^{-13}$ \unit{\meter^{-2/3}} or more.
On the other hand, the effect of turbulence on the fluctuations of the intensity also depends on the propagation distance $L$ and the wavelength. 
In particular, irradiance fluctuations in turbulent environments are usually characterized by the Rytov variance~\cite{Andrews2005}
\begin{equation}
  \sigma_R^2(L)=1.23 C_n^2 k^{7/6}L^{11/6},
  \label{eq:rytov}
\end{equation}
which is actually the scintillation index of a plane wave in Kolmogorov turbulence. Specifically, if $\sigma_R^2(L)\ll1$, $\sigma_R^2(L)\sim1$, and $\sigma_R^2(L)\gg1$ the intensity fluctuations are classified as weak, moderate, and strong, respectively.

In our simulations, we use a split step Fourier method to derive the beam dynamics and uncorrelated (Markov) phase screens to simulate the random refractive index fluctuations. The computational window consists of $N\times N$ points with resolution $\delta_x=\delta_y=1$ \unit{\milli\meter}, to properly sample high-$k$ beam components. We use a numerical grid of $N=1024$ points, resulting in a computational window of approximately $1$ \unit{\meter\squared} cross-section. To absorb the energy that possibly spreads beyond the extent of the grid at the onset of turbulence, we select absorptive boundaries and apply a super-Gaussian filter function~\cite{Schmidt2010}.

The propagation distance of the simulations is $L=(\Delta z)N_z$, where $\Delta z$ is the distance between successive phase screens and $N_z$ is the number of phase screens. The two-dimensional phase screens are generated using the sparse uniform method~\cite{Charnotskii2020}.
In our simulations we use 500 spectral components for each screen. The Rytov variance between two successive phase screens $\sigma_R^2(\Delta z)$
is selected to be smaller than $0.1$~\cite{Xiao2009} for all simulated examples resulting in a total of $N_z=\numrange{10}{20}$ phase screens.
For propagation along the Earth's surface, turbulence can be considered homogeneous and isotropic and, hence, for the refractive index fluctuations, we use the modified von K\'arm\'an power spectral density model
\begin{equation}
\Phi_n(\kappa) = 0.033 C_n^2 \frac{\exp{(-\kappa^2/\kappa_m^2)}}{(\kappa^2 + \kappa_0^2)^{11/6}},
\end{equation}
where $\kappa_m=5.92 / l_0$, $\kappa_0=2\pi / L_0$ and $l_0$, $L_0$ are the inner and outer scales of the inertial subrange, respectively. Specifically, we select $l_0=5$ \unit{\milli\meter} and $L_0=10$ \unit{\meter}, which are typical values for propagation close to the surface of the earth~\cite{Andrews2005}.

\begin{figure}[t!]
  \centering
  \includegraphics[width=\linewidth]{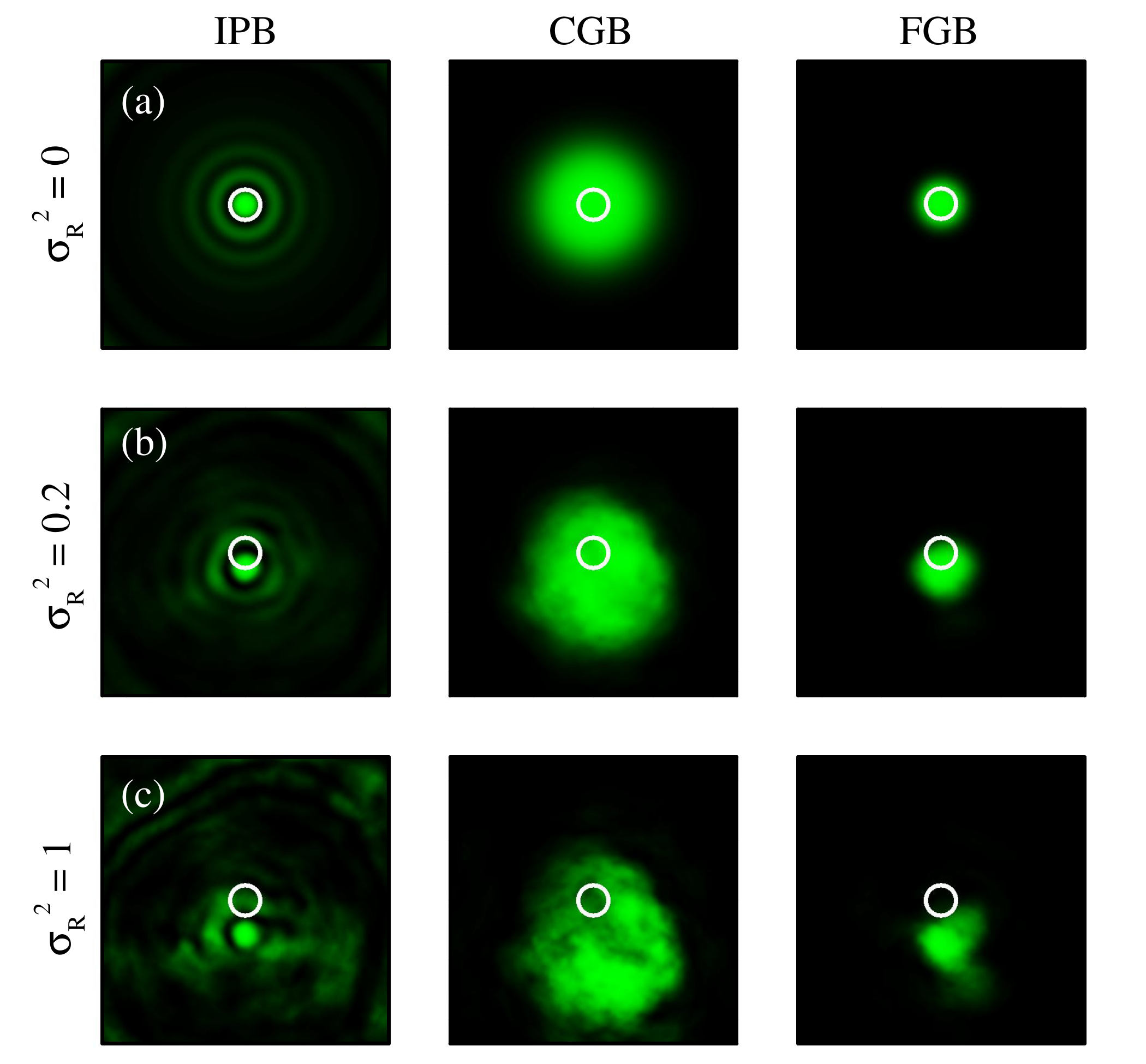}
  \caption{Intensity snapshots of IPB, CGB, and FGB profiles, after propagation of $L=6$ \unit{\kilo\meter} in a medium that leads to (a) no fluctuations, $\sigma_R^2=0$, (b) weak fluctuations, $\sigma_R^2=0.2$, and (c)  moderate fluctuations, $\sigma_R^2=1$. The white circle marks the $1$ \unit{inch} receiver aperture area. The beam parameters are the same as in Fig.~\ref{fig:01}.}
  \label{fig:03}
\end{figure}

To ensure a fair comparison, we set a common receiver distance of $L=6$ \unit{\kilo\meter}, which is the propagation distance at which we desire to minimize the scintillation index of each beam. As a tuning parameter, we use the maximum ray propagation distance, $z_\mathrm{max}$, for IPBs, PBs, and BBs, and the beam radius, $w_0$, for the Gaussian beams. The focal distance $f$ for FGB is selected to be equal to the receiver distance.

In Fig.~\ref{fig:02} we present the scintillation index for each beam calculated under different turbulence strengths.
We take ensemble average over $1000$ realizations, using the same screens for all types of the examined beams, and a $D_r=1$ \unit{inch} receiver aperture. In all our numerical calculations, for the computation of the scintillation index, the intensity is averaged on the receiver aperture. For weak fluctuations, ($C_n^2 = 10^{-16}$ \unit{\meter^{-2/3}}, $\sigma_R^2=0.185$) the scintillation index of IPBs and BBs decreases with $z_\mathrm{max}$ [Fig.~\ref{fig:02}(a)-(b)]. On the other hand, for PBs there is an optimum value between $z_\mathrm{max}=\numrange{8}{10}$ \unit{\kilo\meter} [Fig.~\ref{fig:02}(c)]. We select a common value of $z_\mathrm{max}=8$ \unit{\kilo\meter} which is also a good option for intermediate-strong irradiance fluctuations ($C_n^2 = 10^{-15}$ \unit{\meter^{-2/3}}, $\sigma_R^2=1.85$) to very strong fluctuations ($C_n^2 = 10^{-14}$ \unit{\meter^{-2/3}},  $\sigma_R^2=18.5$). 
For CGBs and FGB [Fig.~\ref{fig:02}(d)] with $C_n^2 = 10^{-16}$ \unit{\meter^{-2/3}}, $\sigma_I^2$ is minimized for $w_0=20$ and $28$ \unit{\milli\meter}, respectively~\cite{Miller1994, Miller1995, Andrews2005}.
Increasing  $C_n^2$ to $10^{-15},10^{-14}$ \unit{\meter^{-2/3}}, the scintillation index of CGBs and FGBs takes similar values. For the first (latter) case $\sigma_I^2$ increases (decreases) with $w_0$. We optimize both regimes by selecting $w_0=20,50$ \unit{\milli\meter}, for CGBs and FGBs, respectively. 
We see a general trend of IPB, BB, and PB to have reduced scintillation index as compared to Gaussian beams especially as the strength of turbulence increases. Overall, IPBs outperform all the other families of examined beams.

To understand the physical processes behind the observed performance, in Fig.~\ref{fig:03} we plot single realizations of each beam after $6$ \unit{\kilo\meter} of propagation. We use the same phase screen set in order to simulate identical turbulence conditions and vary the turbulence strength. In the weak fluctuation regime [Fig.~\ref{fig:03}(b), $\sigma_R^2=0.2$], the CGB at the receiver exhibits small scale scintillations (a speckled intensity profile). On the other hand, FGBs and IPBs preserve their structure, and are more prone to beam wander. For stronger fluctuations ($\sigma_R^2=1$) these effects are intensified: In the realization shown in Fig.~\ref{fig:03}(c) both CGB and FGB exhibit significant wander resulting in a large part of the beam being outside the receiver aperture. The transverse displacement of the IPB is relatively smaller than the FGB. Note that the IPB is able to maintain its shape (including the first side lobe) which is also important in reducing the scintillation index.

\begin{figure}[t!]
  \centering
  \includegraphics[width=\linewidth]{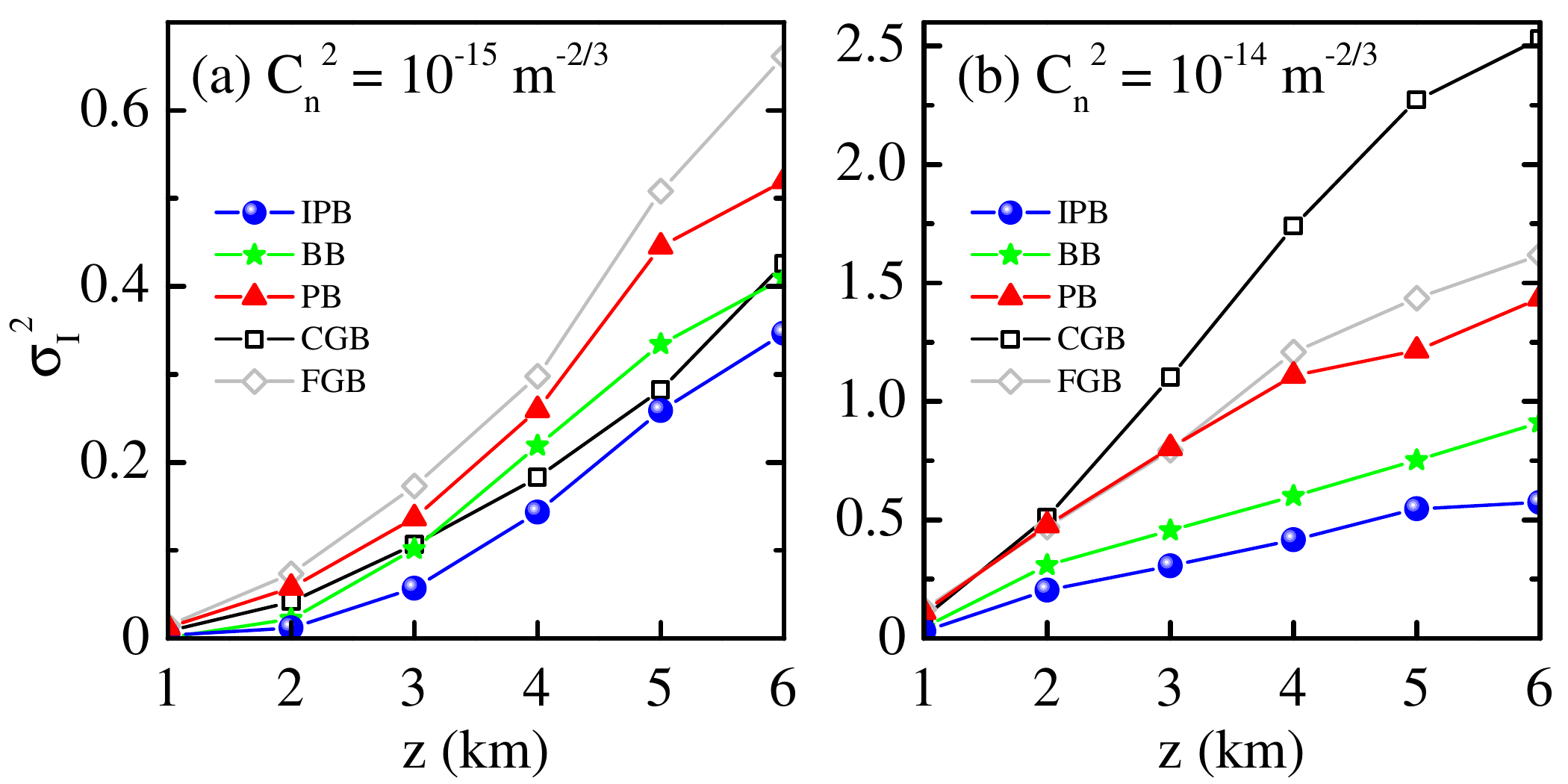}
  \caption{Scintillation index vs propagation distance for beams propagating in (a) moderate turbulence ($C_n^2=10^{-15}$ \unit{\meter^{-2/3}}) and (b) strong turbulence ($C_n^2=10^{-14}$ \unit{\meter^{-2/3}}). All beams are selected to have optimum scintillation at the target distance of $L=6$ \unit{\kilo\meter}.}
  \label{fig:04}
\end{figure}

A beam optimized for all atmospheric conditions and possible propagation distances would be certainly ideal for applications. However, the multi-parametric nature of the problem does not allow for such a solution. The numerical study in Fig.~\ref{fig:02} showed that the minimization of scintillation depends on several parameters, including the beam characteristics, the target distance, as well as the turbulence conditions. In general, it is not practical or possible to change the beam design depending on these parameters. Here, based on the designs of the previous section, we would like to explore how
$\sigma_I^2$
varies when the receiver is placed at distances smaller than the target distance. 

To investigate this possibility, we calculate the scintillation index between $\numrange{1}{6}$ \unit{\kilo\meter} with $1$ \unit{\kilo\meter} steps, for all the classes of beams examined in the previous section (target distance of $L=6$ \unit{\kilo\meter}). The results are presented in Fig.~\ref{fig:04}. Note that for strong turbulence, the structured light beams with $\gamma=0.4,1,1.4$ outperform the Gaussian beams (both collimated and focused).
Importantly, for both moderate [Fig.~\ref{fig:04}(a)] and strong turbulence [Fig.~\ref{fig:04}(b)] the IPB always has the lowest value of the scintillation index independently of where the receiver is placed. For moderate turbulence they are followed mainly by the CGBs and for strong turbulence by the BBs. The differences in the performance between IPBs and the rest of the beams are more prominent as the strength of turbulence is further increased.

In this work, we have focused on the most important measure for the performance of a beam in free-space optical communications, which is the normalized mean intensity variance (scintillation index). In Supplemental document 1, we also examine the power delivery properties of the same classes of beams, and discuss possible ways to further optimize their capabilities.

In conclusion, we have proposed a new class of structured light beams, with a Bessel-like transverse profile and width that gradually increases during propagation, namely an inverted pin beam. Using a thorough numerical optimization procedure, we have compared these beams with four other classes of Gaussian (collimated and focused), Bessel, and pin beams. We have found that inverted pin beams have better performance in terms of their scintillation index in environments with moderate and strong atmospheric turbulence. In addition, the same inverted pin beam design can be successfully applied to a variety of different turbulent conditions and propagation distances.
We expect that such beams might be particularly useful in applications related to laser beam propagation in turbulent media, such as free-space optical communications, as well as in optical imaging.

\begin{backmatter}
  
  \bmsection{Funding} This work was supported by Huawei Technologies Sweden AB.
  
  \bmsection{Acknowledgments} The authors would like to thank Ulrik Imberg from Huawei for his assistance in this project and Domenico Bongiovanni for discussions.

  \bmsection{Disclosures} The authors declare no conflicts of interest.

  \bmsection{Data Availability Statement} Data underlying the results presented in this paper are not publicly available at this time but may be obtained from the authors upon reasonable request.

  \bmsection{Supplemental document} See Supplement 1 for supporting content. 

\end{backmatter}

\cleardoublepage

\renewcommand\refname{FULL REFERENCES}

\newpage



\section*{Supplementary material}

# Inverted pin beams for robust long-range propagation through atmospheric turbulence: supplemental document

## 1. Power delivery

The most important parameter for optical communication systems is the scintillation index. For example, the signal-to-noise ratio and the bit error rate are directly related to the scintillation index which is the normalized variance of the mean intensity [1]. An additional parameter that should be considered, is the power delivery capability of a beam, i.e., the amount of power captured by the receiver divided by the total power on the transmitter. We have computed the power delivery of all the beam profiles shown in Fig. 2 of the main text for the same parameters. These results are shown in Supplemental Figure 1, where we depict ensemble average values. The main outcome is that, under the same amount of turbulence, the fraction of power on the receiver is of the same order between all these classes of beams. Thus, there is no fundamental drawback from using any of these beams.

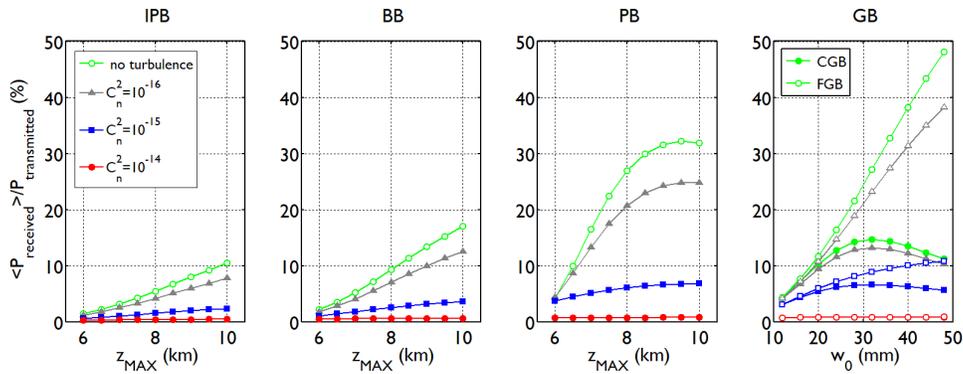

**Supplemental Figure 1.** Fraction (%) of the power on the receiver having 1 inch aperture over the power on the 20 cm transmitter aperture. The parameters and the horizontal axes are the same as in Fig. 2 of the main text.

Additionally, it is desirable that the beam maintains a relatively high intensity spot on the receiver. In this respect, we have measured the ensemble average fraction of on-axis beam intensity divided with the total power on the receiver. The results are shown in Supplemental Figure 2 where, again, the comparison involves the beam profiles of Fig. 2 of the main text.

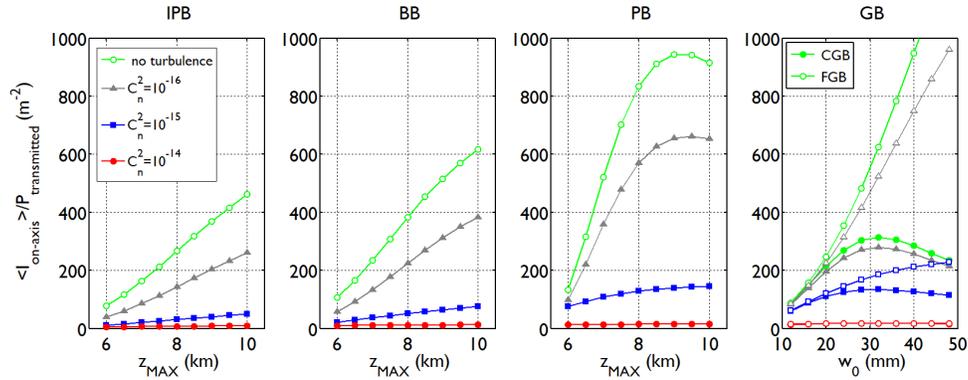

**Supplemental Figure 2.** Fraction (%) of the on-axis intensity on the receiver over the power on the 20 cm transmitter aperture. The parameters and the horizontal axes are the same as in Fig. 2 of the main text.

## 2. Optimization of pin beams for power delivery

The beam profile of Eq. (7) is important from the theoretical point of view. It shows that pin beams can be considered as a natural generalization of Bessel beams. However, in terms of applications it is far simpler to generate beams via phase modulation as given by Eq. (1) of the main text. Importantly, the profile of Eq. (1) contains only the rays that converge towards the axis during propagation. Since the diverging rays contain 50% of the total power of the beam, by using the profile of Eq. (1) will result to doubling the intensity and, as a result, the power delivery capabilities of pin type of beams. Thus, the power and intensity delivery of the first three columns of both figures of this supplement will see an increase by a factor of 2.



\end{document}